\documentclass{JINST} % 

\title{LFI 30 and 44 GHz receivers Back-end Modules}

\author{ 
E.~Artal$^a$\thanks{Corresponding
author.}, B.~Aja$^a$, M.~L.~de~la~Fuente$^a$, J.~P.~Pascual$^a$, A.~Mediavilla$^a$, E.~Martinez-Gonzalez~$^b$ ,
L.~Pradell$^c$, P.~de~Paco$^c$,
M.~Bara$^d$, E.~Blanco$^d$, E.~Garc\'ia$^d$,
R.~Davis$^e$, D.~Kettle~$^e$, N.~Roddis$^e$, A.~Wilkinson$^e$,
M.~Bersanelli$^f$, A.~Mennella$^f$, M.~Tomasi$^f$, R.~C.~Butler$^g$, F.~Cuttaia $^g$, N.~Mandolesi$^g$ and L.~Stringhetti$^g$\\
\llap{$^a$}Dpt. Ingenier\'ia de Comunicaciones, Universidad de Cantabria,\\ Plaza de la Ciencia, 39005 Santander Spain\\     
\llap{$^b$}Instituto de F\'isica de Cantabria, CSIC-Universidad de Cantabria,\\Avda. de los Castros s/n, 39005 Santander, Spain\\
\llap{$^c$}Departament de Teor\'ia del Senyal i Comunicacions, Universitat Polit\'ecnica de Catalunya,\\Barcelona, Spain\\
\llap{$^d$}Mier Comunicaciones S.A.\\La Garriga, Barcelona, Spain\\
\llap{$^e$}Jodrell Bank Centre for Astrophysics, University of Manchester,\\United Kingdom\\
\llap{$^f$}Universit\'a degli studi di Milano, Department of Physics\\Via Celoria 16, Milano, Italy\\
\llap{$^g$}INAF/IASF - Bologna\\Via P. Gobetti 101, I-40129, Bologna, Italy\\

 E-mail: \email{artale@unican.es}}
 
 \abstract{ The 30 and 44 GHz Back End Modules (BEM) for the Planck Low Frequency Instrument 
  are broadband receivers (20\% relative bandwidth) working at room temperature. 
  The signals coming from the Front End Module are amplified, band pass filtered and finally converted to DC by a detector diode. Each receiver has two identical branches following the differential scheme of the Planck radiometers. The BEM design is based on MMIC Low Noise Amplifiers using GaAs P-HEMT devices, microstrip filters and Schottky diode detectors. Their manufacturing development has included elegant breadboard prototypes and finally qualification and flight model units. Electrical, mechanical and environmental tests were carried out for the characterization and verification of the manufactured BEMs.
A description of the 30 and 44 GHz Back End Modules of Planck-LFI
radiometers is given, with details of the tests done to determine their
electrical and environmental performances.
The electrical performances of the 30 and 44 GHz  Back End Modules:
frequency response, effective bandwidth, equivalent noise temperature, 1/f
noise and linearity are presented.}


\begin{document}

\section{Introduction}
The Low Frequency Instrument (LFI) is one of the two instruments of the ESA Planck satellite. The objective of the satellite is to achieve precision maps of the Cosmic Microwave Background with a very good sensitivity to observe sky temperature anisotropies \cite{mandolesi09}. LFI contains very high sensitive receivers based on cryogenic radiometers with their Front-End Modules (FEM) operating at 20 $K$ \cite{davis09}. LFI covers three adjacent frequency bands centred at 30, 44 and 70 $GHz$, with a fractional bandwidth of 20\%, and uses cryogenic very low noise amplifiers with Indium  Phosphide (InP) High Electron Mobility Transistors (HEMT) in the FEM. The Front-End Modules are physically and thermally separated from the Back-End Modules by over a meter of copper and stainless steel rectangular waveguide. The signals originating in the FEM are then further amplified and finally detected in the BEM that operate at 300 $K$.

\begin{figure}[t]
\centering
\includegraphics[width=.8\textwidth]{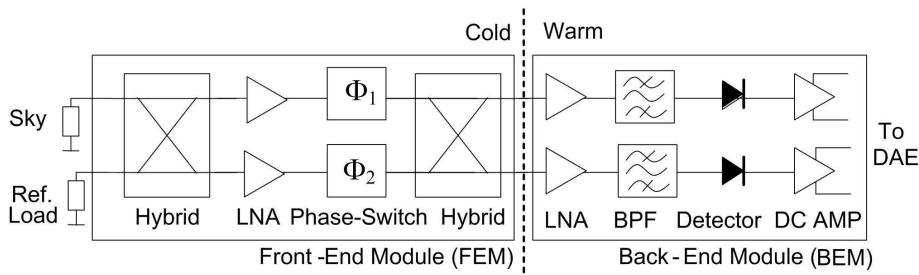}
\caption{Planck Mission radiometer scheme.}
\label{fig:fig1}
\end{figure}

Planck-LFI radiometers are based on a differential scheme \cite{blum59}, \cite{bersanelli02}, \cite{bersanelli09}, shown in Figure \ref{fig:fig1}. Signals from the sky and from a 4 $K$ reference load are combined using a $(180^{\;\circ})$ hybrid. Both signals are amplified by Low Noise Amplifiers (LNA) in the Front-End Module. Two identical phase switches introduce a differential $(180^{\;\circ})$ phase shift in one branch with relation to the other. The second $(180^{\;\circ})$ hybrid delivers a signal proportional to sky temperature in one output, and a signal proportional to the reference load temperature in the other.

%\begin{figure}
%		\centering
%		\includegraphics{./fig1.eps}
%		\caption{Planck Mission radiometer scheme.}
%	\label{fig:fig1}
%\end{figure}

The balanced structure of the FEM permits to obtain two signals, sky and reference, contaminated by the same amount of noise because both have passed through the same LNA and phase switches. By differencing signals across the two outputs using post-processing, the noise can be cancelled. Phase switching at 4 $kHz$ is done also to remove the 1/f noise of the BEM, because it does not have a balanced structure like the FEM. This BEM noise is noticeable at low frequencies, at some hundreds of $Hz$. Noise cancellation can only be perfect if the two FEM branches are identical and the balance is perfect as well. In practice a little unbalance is tolerated and it was tested as a leakage from one input to the theoretically isolated output.

Next sections describe in detail the 30 and 44 $GHz$ BEM units of Planck-LFI, from their design principles and individual subsystems performance, to the final Flight Model units which are integrated in the Planck satellite. Prototype and Elegant Breadboard units are presented and described showing their main characteristics. Electrical and environmental tests have been performed in all the Qualification and Flight Model units delivered to the Planck satellite Project System Team.

\section{Design of the Back-End Modules}

In this section, we will describe all the fundamentals of each RF component designed and their individual performance before integration.

\subsection {Low Noise Amplifiers}

\begin{figure}[t]
\centering
\includegraphics[width=1\textwidth]{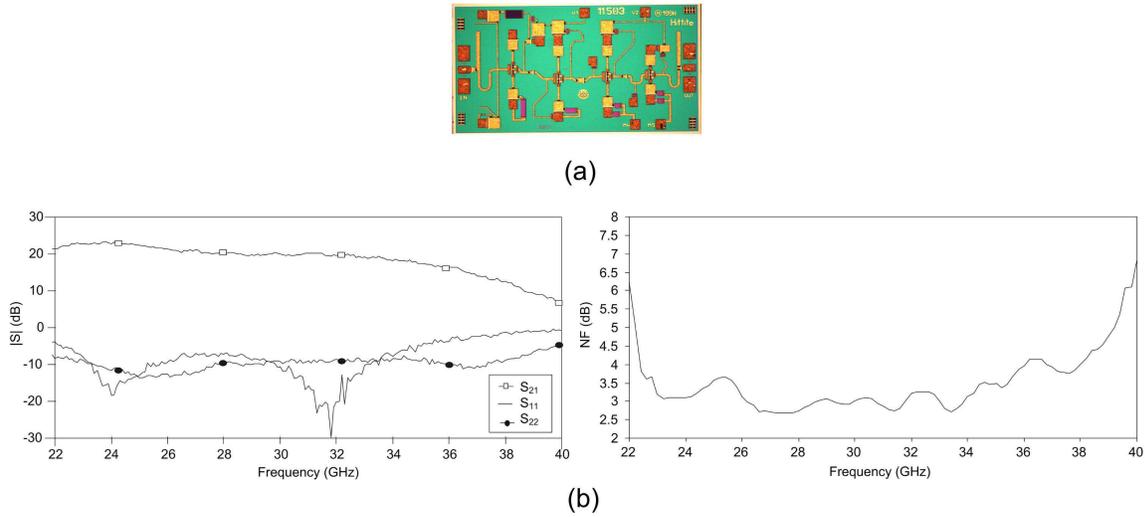}
\caption{(a) Ka-band LNA MMIC (b) MMIC performance.}
\label{fig:fig2}
\end{figure}

Each Low Noise Amplifier (LNA) consists of two cascaded amplifiers in order to have enough gain and to provide a signal in the square law region of the diode detector. Monolithic Microwave Integrated Circuits (MMIC) on GaAs technology have been chosen at 30 and $44$ $GHz$. The MMIC at Ka-band are commercial circuits, model HMC263 from Hittite. They have four stages of Pseudomorphic-High Electron Mobility Transistors (PHEMT), with an operating bandwidth from 24 $GHz$ to 36 $GHz$, and offer 23 $dB$ of gain and 3 $dB$ of noise figure from a self-biasing supply of $+$3 $Volt$, 52 $mA$. A picture of the MMIC at Ka-band, its noise figure, gain and input and output return loss are shown in Figure \ref{fig:fig2}.

\begin{figure}[t]
\centering
\includegraphics[width=.6\textwidth]{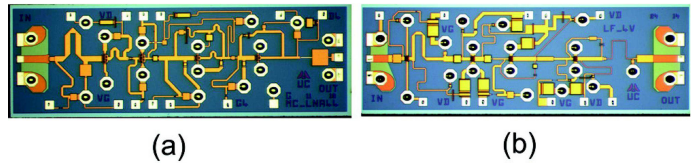}
\caption{Q-band LNAs, (a) Normally ON; (b) Normally OFF.}
\label{fig:fig3}
\end{figure}

\begin{figure}[t]
\centering
\includegraphics[width=1\textwidth]{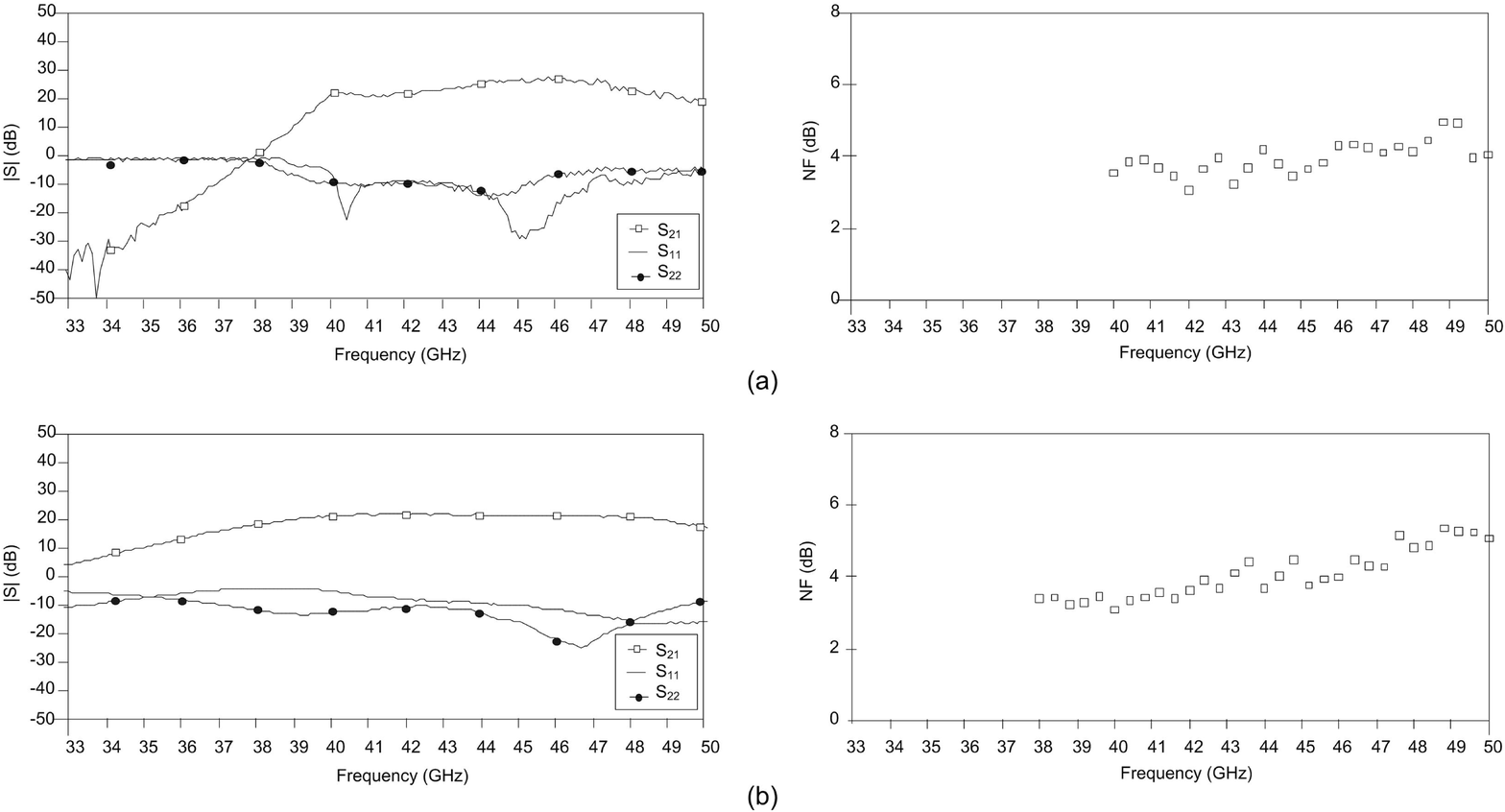}
\caption{Q-band LNAs on wafer performance, (a) Normally ON; (b) Normally OFF.}
\label{fig:fig4}
\end{figure}

At Q-band two custom designed MMICs have been assembled \cite{aja02}. They have been manufactured with the process ED02AH from OMMIC, which employs a 0.2 $\mu m$ gate length PHEMT on GaAs. The first MMIC has four stages of depletion mode transistors (Normally ON: N-ON) with gate widths of 4x15 $\mu m$ and 6x15 $\mu m$, shown in Figure \ref{fig:fig3} $($a$)$. The second MMIC, in Figure \ref{fig:fig3} $($b$)$, has four-stages of enhancement mode transistors (Normally-OFF: N-OFF) with gate widths of 6x15 $\mu m$. The N-ON LNA preceding the N-OFF, was found to be the best in terms of input and output matching and noise performance. The main requirements of the two LNAs are to provide low noise and low power consumption with enough gain. Using as first amplifier an LNA, based on N-ON PHEMT transistors, a good noise performance was obtained with a power consumption of only 90 $mW$. The second LNA, with N-OFF PHEMT transistors, was added to increase the gain with a very low impact in the noise figure and a very low power consumption of 32 $mW$.
All the circuits were measured on wafer using a coplanar probe station. Noise figure, return loss and associated gain for the Q-band depletion and enhancement transistor LNAs are plotted in Figure \ref{fig:fig4}.


\subsection{Band Pass Filters}

\begin{figure}[t]
\centering
\includegraphics[width=.4\textwidth]{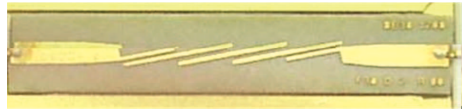}
\caption{Microstrip bandpass filter.}
\label{fig:fig5}
\end{figure}

\begin{figure}[t]
\centering
\includegraphics[width=1\textwidth]{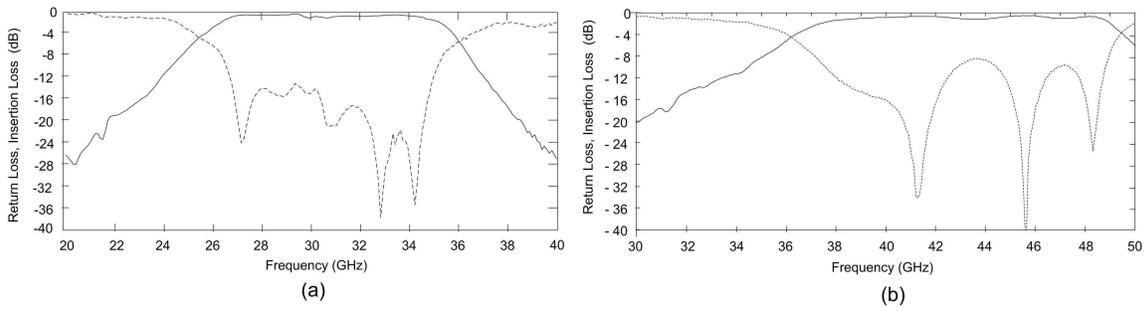}
\caption{Band-pass filter performance, (a) 30 $GHz$; (b) 44 $GHz$.}
\label{fig:fig6}
\end{figure}

A band pass filter was used to define an effective bandwidth of 20\% and to reject undesired signals out of the band of interest. Low bandpass losses, more than 10 dB out of band losses and small size were considered the main objectives to fulfil. The filter was based on microstrip coupled line structure that was chosen because it provides inherently good band pass characteristics. The design is a three-order Chebyshev resonator filter. Electrical models of coupled lines and an electromagnetic simulator was used in the design phase. The design method was based on the classic prototype filter tables provided by \cite{matthei64}, but a design methodology has been developed to achieve a predictable frequency response in microstrip filters using commercial CAD software. After a careful evaluation of the validity of the CAD models, comparing simulated and accurately measured results, the design was restricted to microstrip elements than could be well characterized \cite{detratti02}. The selection of the substrate became critical due to the gaps and widths of the microstrip lines because it sets the line-etching precision required and the minimum losses achievable. Microstrip lines were made on Duroid 6002 substrate with 10 mils thickness and dielectric constant of 2.92. Several units of microstrip band pass filters were fabricated and tested. Typical test results for the 30 $GHz$ filter are insertion losses lower than 0.84 $dB$ in the band, and for the 44 $GHz$ filter, insertion losses better than 1.5 $dB$ and return losses better than 10 $dB$ across the operating bandwidth.
A photograph of a 30 $GHz$ filter is shown in Figure \ref{fig:fig5}. The response of the filters when they have been measured with coplanar to microstrip transitions using a coplanar probe station is depicted in Figure \ref{fig:fig6}.


\subsection{Diode Detectors}

\begin{figure}[t]
\centering
\includegraphics[width=.6\textwidth]{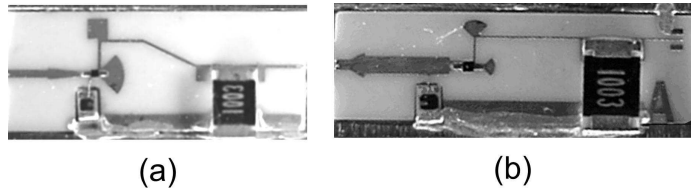}
\caption{Diode detectors, (a) 30 $GHz$ ; (b) 44 $GHz$.}
\label{fig:fig7}
\end{figure}

\begin{figure}[t]
\centering
\includegraphics[width=1\textwidth]{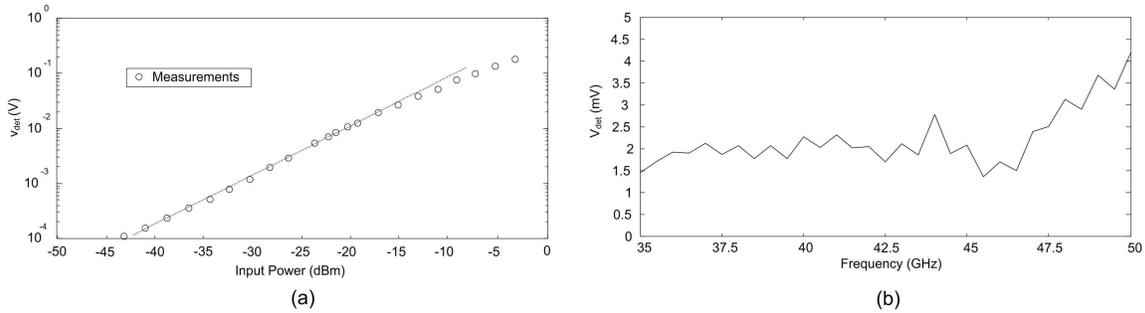}
\caption{(a) Ka-band detector sensitivity at 30 GHz ; (b) Q-band detector rectification efficiency at 44 $GHz$ for -30 $dBm$ of input power.}
\label{fig:fig8}
\end{figure}

The diode for the detector was a GaAs planar doped barrier Schottky diode. The specific model chosen was a zero-bias beam-lead diode HSCH-9161 from Agilent Technologies. This diode has suitable characteristics to be used at microwave frequencies. Among the main specifications in the detector design are the input matching, the sensitivity and the tangential sensitivity. The diode equivalent circuit is not a good match to 50 $Ohm$, so it was necessary to synthesize a network that would transform it to something close with an input matching network. Thus the detector is composed of a hybrid reactive/passive matching network, and the Schottky diode. Both detectors, for 30 $GHz$ and 44 $GHz$, were mounted with a coplanar-to-microstrip transition to make on-wafer tests as a previous step to the BEM integration. The practical implementation is performed with transmission lines printed on a standard dielectric substrate, Alumina with 10 $mils$ thickness and dielectric constant of 9.9. A view of the detectors is shown in Figure \ref{fig:fig7}. The output voltage sensitivity has been measured at 30 $GHz$ for the Ka band detector and it is shown in Figure \ref{fig:fig8} $($a$)$. Figure \ref{fig:fig8} $($b$)$$ $ depicts the rectification efficiency of the Q-band detector at the frequency of 44 $GHz$ for $-$30 $dBm$ of input power.


\subsection{DC Amplifier}

\begin{figure}[t]
\centering
\includegraphics[width=.6\textwidth]{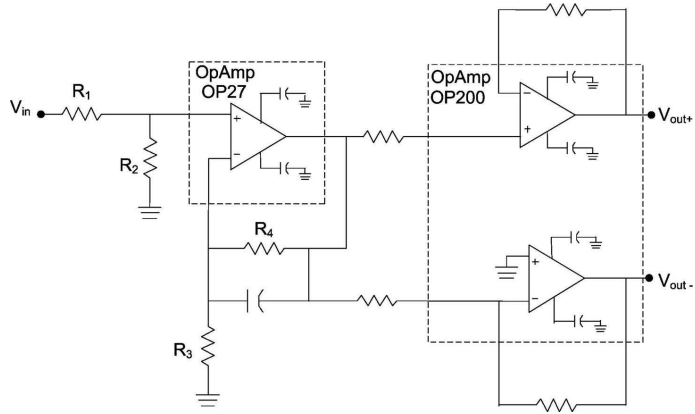}
\caption{DC-amplifier schematic.}
\label{fig:fig9}
\end{figure}

The detector diode output is connected to a low noise DC-amplifier, with an adequate voltage gain to have the required detected signal level for the data acquisition electronic module. A schematic of the DC-amplifier is shown in Figure \ref{fig:fig9}. The first stage has an OP27 precision operational amplifier that combines low offset and drift characteristics and low noise, making it ideal for precision instrumentation applications and accurate amplification of a low-level signal. A second balanced stage, implemented with an OP200, provides a balanced and bipolar output. DC amplifier total power consumption with a high impedance load is 37 $mW$.

The amplifier gain is given by the ratio between the differential output voltage $V_{out}$, and $V_{in}$ (input voltage from the detector), where $V_{out}$ is given by: 

\begin{equation}
V_{out}=V_{out+}-V_{out-}=2\cdot \frac{R_{2}}{R_{1}+R_{2}}\cdot\left[1+\frac{R_{4}}{R_{3}}\right]\cdot V_{in}	
\end{equation}

The detector resistive load is $R_{1}+R_{2}$. In order not to affect the detector RF response, the condition $R_{1}+R_{2}$ $\geq 50$ $kOhm$ was fulfilled.

\begin{figure}[t]
\centering
\includegraphics[width=.6\textwidth]{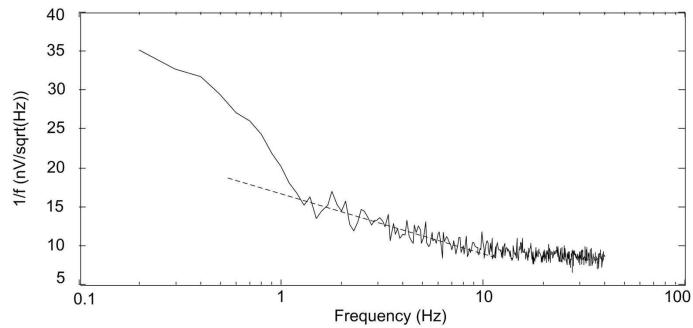}
\caption{ DC amplifier low frequency noise, referred to the amplifier input.}
\label{fig:fig10}
\end{figure}

Since the phase switch frequency rate in the receiver is 4096 $Hz$, the output signal has to provide a video bandwidth of at least 50 $kHz$, which means that the output signal contains more than ten harmonics in order not to degrade the information.
The DC amplifier measured gain-bandwidth product was 5.7 $MHz$. This operational amplifier gain bandwidth product has been taken into account, and the maximum achievable balanced gain without loosing output bandwidth was 100. A voltage gain of 50 is due to the OP27 and a further of 2 due to unbalanced to balanced conversion of the OP200 with unit individual gain. Another constraint of this DC amplifier is to provide an output voltage in a window between 0.2 $Volt$ and 0.8 $Volt$, where the data acquisition electronics (DAE) works properly. The designed DC amplifier was adjusted for each channel, taking into account small RF gain differences, in order to have the output DC voltage inside the window and to achieve the output bandwidth requirement. The resulting bandwidth was 163 $kHz$.
Figure \ref{fig:fig10} shows the DC amplifier measured low-frequency noise referred to the amplifier input. The white noise voltage level is $8$ $nV/\sqrt{Hz}$ $(=-162$ $dBV/\sqrt{Hz})$, and the flicker noise knee-frequency, defined as the point where the noise voltage is $\sqrt{2}$ times the white noise voltage, is 2.8 $Hz$.

\section{Manufacturing}
\subsection{Elegant breadboard prototype}
\begin{figure}[t]
\centering
\includegraphics[width=.7\textwidth]{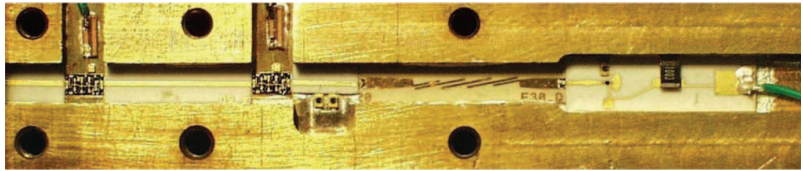}
\caption{General view of the EBB 30 $GHz$ BEM branch with LNA, band pass filter and detector.}
\label{fig:fig11}
\end{figure}

\begin{figure}[t]
\centering
\includegraphics[width=.6\textwidth]{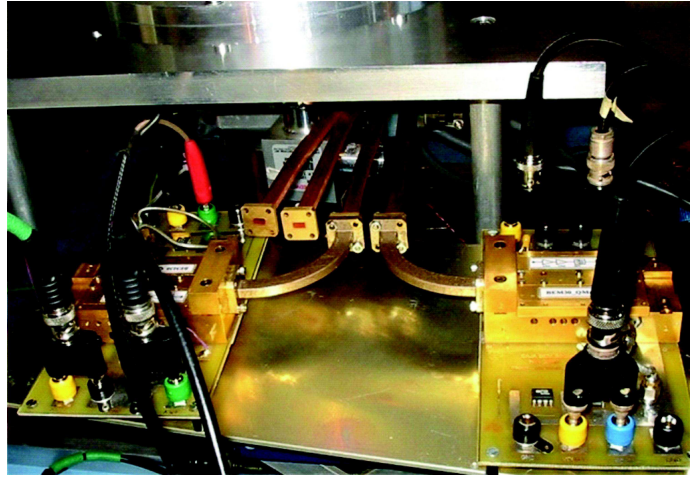}
\caption{The EBB 30 GHz BEM branches connected to the Prototype Demonstrator FEM in Jodrell Bank Observatory.}
\label{fig:fig12}
\end{figure}

The objective of the Elegant Breadboard (EBB) prototypes was to experimentally demonstrate the radiometer concept, by integrating a fully representative unit in the laboratory. An EBB demonstrator must have the same electrical functionality as the final flight unit. The internal architecture, electrical scheme, and the electronic components were the same as for flight. The electronics components do not need to be space qualified. The size and weight of the EBB demonstrator can be different from the flight unit. To make easier the integration with the EBB of the Front End Module, the EBB version of the 30 and 44 $GHz$ BEM included only one branch. In fact the EBB version is a quarter of one full BEM. Figure \ref{fig:fig11} shows a photograph of one EBB at 30 $GHz$. This EBB branch contains a Low Noise Amplifier (LNA), a Band Pass Filter (BPF), a Schottky diode detector (DET) and a low frequency amplifier (DC Amp). The central frequency is 30 $GHz$ and the nominal fractional bandwidth is 20\%.  Figure \ref{fig:fig12} shows two EBB BEM units connected to the Prototype Demonstrator (PD) FEM in the laboratory at Jodrell Bank Observatory.


\begin{figure}[t]
\centering
\includegraphics[width=1\textwidth]{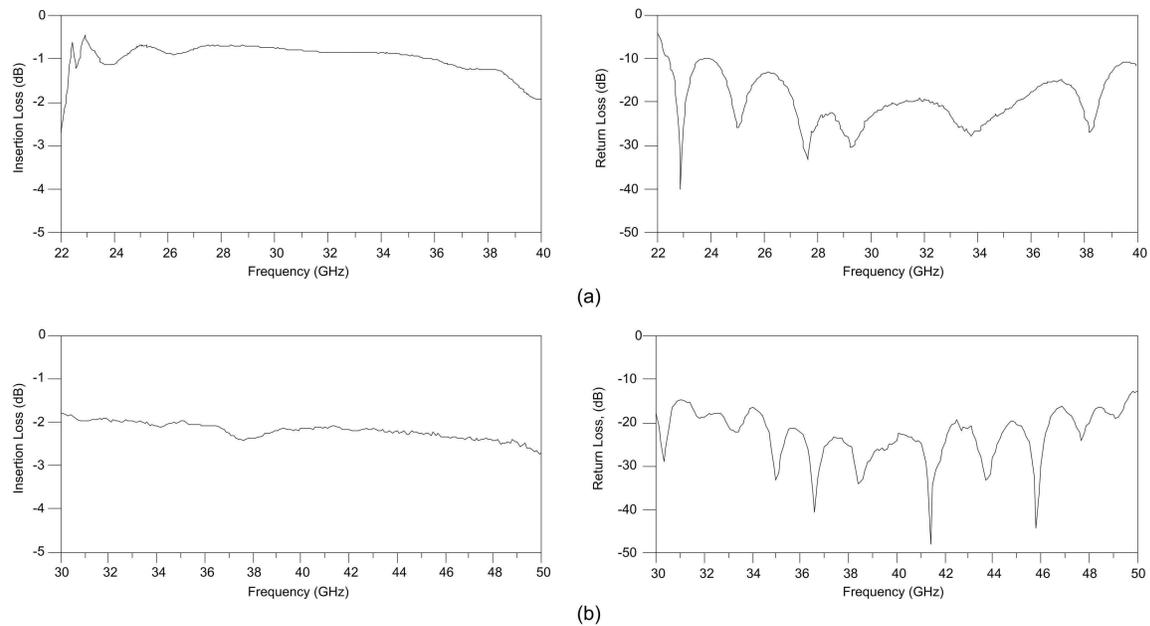}
\caption{Insertion loss and return loss of back to back rectangular waveguide to microstrip transitions through stepped ridge sections.(a) Ka band unit, (b) Q band unit.}
\label{fig:fig13}
\end{figure}

The first functional element of each BEM is a waveguide-to-microstrip transition that was designed using a ridge waveguide \cite{hoefer82}. The input waveguide flange is WR-28 for the 30 $GHz$ BEM and WR-22 for the 44 $GHz$ BEM. The transition is a four sections stepped ridge waveguide to microstrip line transition which provides a broadband performance. This transition has been chosen for its broad bandwidth, low insertion loss, and repeatable performance. Figure \ref{fig:fig13} shows the experimental results of back to back rectangular waveguide to microstrip transitions using ridge waveguide transformer. Microstrip line losses are included in the result. The length of microstrip 50 Ohm line, on Alumina substrate 10 mils thickness, was 10 mm in both units (Ka band and Q band). The estimated insertion loss of one single ridge waveguide transition is lower than 0.2 dB.

\subsection{Qualification and Flight Model units}
\begin{figure}[t]
\centering
\includegraphics[width=.5\textwidth]{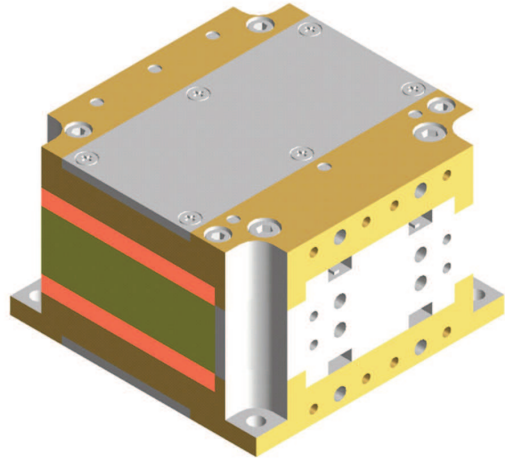}
\caption{Back End module external view. }
\label{fig:fig14}
\end{figure}

The final BEM mechanical configuration for the Qualification Model and Flight Model has four RF branches, providing signal amplification and detection for two complete radiometers. Mechanical design had been carried out by Mier Comunicaciones S.A. within an allowed envelope of 70 x 60 x 39 $mm^{3}$ including all the RF and DC circuitry. Mass is 305 $g$ for 30 $GHz$ BEM and 278 $g$ for the 44 $GHz$ BEM. Figure \ref{fig:fig14} shows the external view of the BEM. Internally there are 5 different levels of PCB circuits, from top to bottom, as follows: 

\begin{enumerate}
	\item DC PCB:  It contains voltage regulators for a half BEM (one receiver, two branches)
	\item DC amplifiers PCB: It contains the DC operational amplifiers for a half  BEM (two signal detected outputs)
	\item RF part: It contains two RF branches (one receiver)
	\item DC amplifiers PCB: It contains the DC operational amplifiers for the other half  BEM (two signal detected outputs)
	\item DC PCB:  It contains voltage regulators for the other half BEM (one receiver, two branches)
	
\end{enumerate}

\begin{figure}[t]
\centering
\includegraphics[width=.6\textwidth]{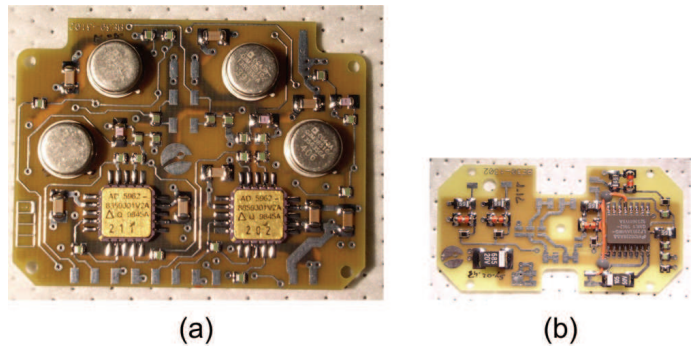}
\caption{(a) DC Amp PCB contains operational amplifiers; (b) DC PCB contains voltage regulators.}
\label{fig:fig15}
\end{figure}

\begin{figure}[t]
\centering
\includegraphics[width=1\textwidth]{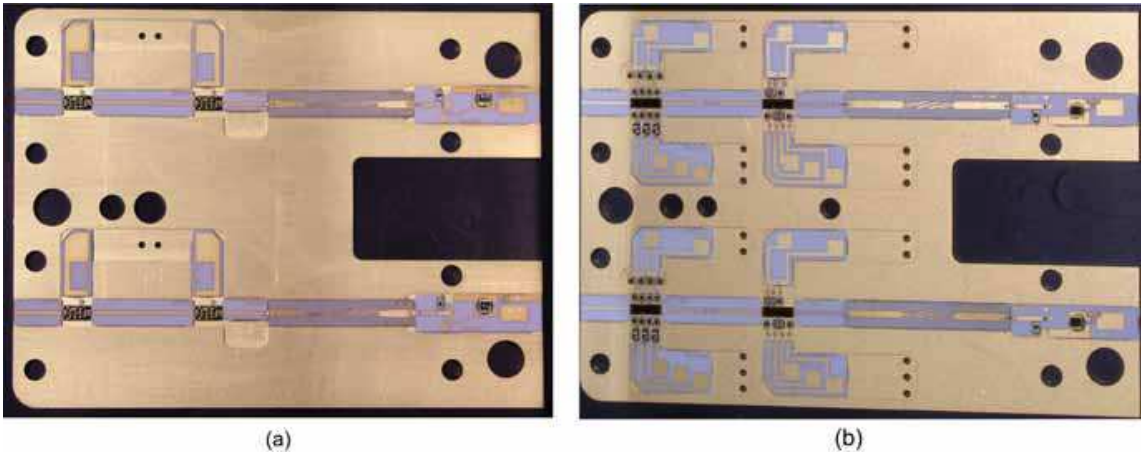}
\caption{RF parts (two branches) in: (a) 30 $GHz$ BEM, (b) 44 $GHz$ BEM.}
\label{fig:fig16}
\end{figure}

Pictures of the DC and RF circuits are in Figure \ref{fig:fig15} and Figure \ref{fig:fig16}.


\begin{figure}[t]
\centering
\includegraphics[width=1\textwidth]{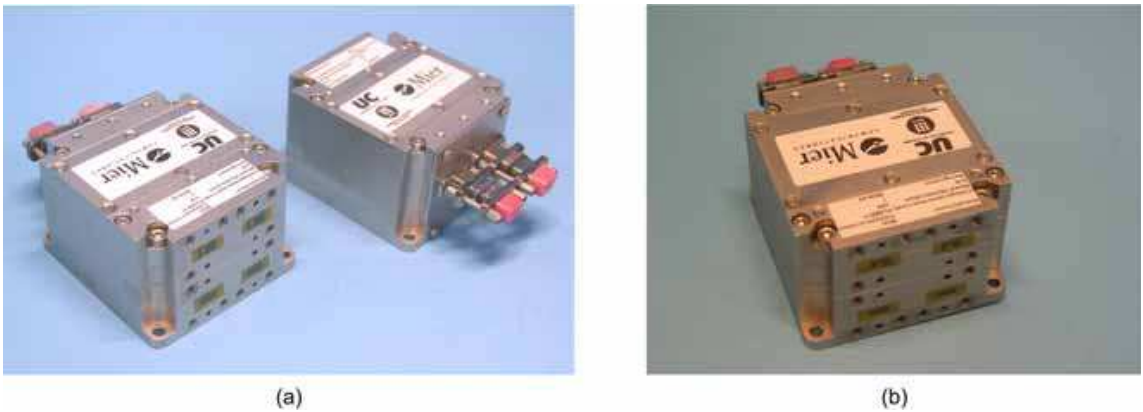}
\caption{(a) FM units of 30 $GHz$ BEM; (b) QM unit of 44 $GHz$ BEM.}
\label{fig:fig17}
\end{figure}

Qualification Model (QM) units have been manufactured using identical electrical and mechanical components as for the Flight Model (FM) units. The QM and FM components have identical quality level and have been space qualified following the same procedures. The only difference between QM and FM units is the different environmental tests done on each case. In particular vibration levels and thermal cycling tests for QM units have been more demanding than for FM units. Summarising: QM units are identical to the FM units, but they are not intended to be installed in the satellite. In order to be ready to deal with limited unexpected component failure in the FM units, before launching the satellite, spare FM units of the 30 and 44 $GHz$ BEM have been also manufactured and tested. Pictures of FM or QM units of 30 and 44 $GHz$ BEM are showed in Figure \ref{fig:fig17}. 

\section{Electrical characterization tests}

A set of basic tests were performed at three different temperatures in the range of possible operating temperature; $T_{low} (-25^{\circ}C)$, $T_{nom} (26^{\circ}C)$ and $T_{high}(48^{\circ} C)$, to verify proper operation of the FM Back-End Modules and fulfillment of the specified electrical requirements. They were assembled and tested in a clean room equipped with a vacuum chamber which enables us to achieve low pressure levels and to vary the base plate and shroud temperature within the acceptance ranges. The principal tests performed were frequency response, equivalent noise temperature, stability and linearity. Test equipment available included vector network analyzers, swept frequency sources, noise sources and low frequency spectrum analyzers.

\subsection{Frequency response and effective bandwidth}
\begin{figure}[t]
\centering
\includegraphics[width=1\textwidth]{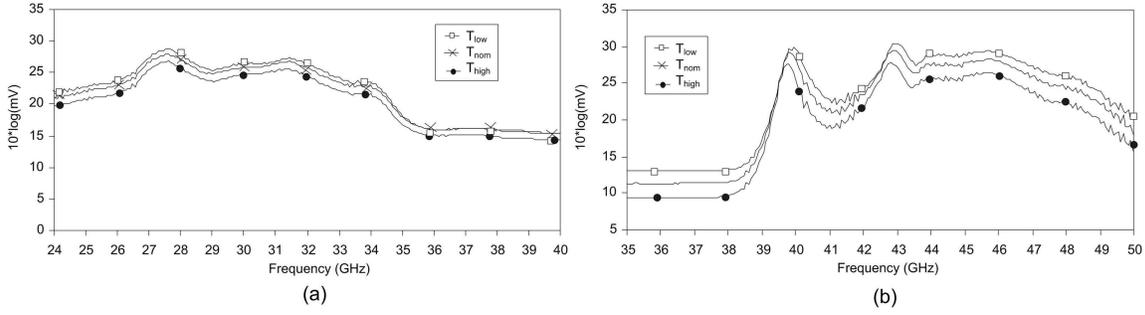}
\caption{RF to DC response at three temperatures for a CW power input of - 60 dBm, (a) BEM 30 $GHz$ Flight Model; (b) BEM 44 $GHz$ Flight Model.}
\label{fig:fig18}
\end{figure}

The frequency response (RF to DC) measurements were done injecting a $CW$ small signal into the waveguide input, for a constant power level. The level of the signal was set to yield a readily measurable response, but not so large as to cause non-linear effects. This test was needed in order to know the bandpass response. It was measured stepping the synthesizer source through 201 frequencies and recording the output voltage for each frequency, with the RF output of the synthesized enabled and disabled. Figure \ref{fig:fig18} shows the RF to DC response for the flight model BEMs at 30 $GHz$ and 44 $GHz$.

The measured results obtained for the RF to DC response are used to calculate the effective bandwidth. The effective bandwidth is defined according to next expresion.

\begin{equation}
BW_{eff}=\frac{\left|\int G(f) df\right|^{2}}{\int\left|G(f)\right|^{2}df}	
\end{equation}

$G(f)$ is the RF power gain of the BEM, including the RF detector response. This $G(f)$ gain is obtained by RF to DC response test. This test is performed with a microwave sweep generator providing a constant input power versus frequency, so the effective bandwidth can be calculated as $($\ref{eq1}$)$, using only the output voltage values taken at discrete frequencies.

\begin{equation}
BW_{eff}=\Delta f \cdot \left(\frac{N}{N+1}\right)\frac{\left(\sum^{N}_{i=1} V_{out}(i)-V_{outoff}\right)^{2}}{\sum^{N}_{i=1} \left(V_{out}(i)-V_{outoff}\right)^{2}}
\label{eq1}	
\end{equation}

where $N$ is the number of frequency points, $\Delta f$ is the frequency step, $V_{out}(i)$ the DC output voltage at each frequency and $V_{outoff}$ is the DC output voltage when the sweep generator is off. $V_{outoff}$ is typically about 25 mV for the 30 GHz BEM and lower values for the 44 GHz BEM.

Table \ref{tab1} shows the values of the effective bandwidth for each flight model BEM at three different temperatures in the range of possible operating temperature.

\begin{table}
\centering
\caption{Effective bandwidth of the BEM Flight models in $GHz$. (One unit of each band is a Flight Spare). Estimated error = $\pm$~1\%.}
\begin{tabular}{c|c|c|c|c}
\hline \hline

$$& Channel $A$ & Channel $B$ & Channel $C$ & Channel $D$ \\
\hline
 
$BEM$&$T_{low}$ $    $ $T_{nom}$ $    $ $T_{high}$&$T_{low}$ $    $ $T_{nom}$ $    $ $T_{high}$&$T_{low}$ $    $ $T_{nom}$ $    $ $T_{high}$&$T_{low}$ $    $ $T_{nom}$ $    $ $T_{high}$\\
\hline
30 $GHz$ FM1 &9.11 $    $ 9.16 $    $ 9.07&9.17 $    $ 9.14 $    $ 9.02&9.34 $    $ 9.36 $    $ 9.27&9.54 $    $ 9.45 $     $ 9.28\\
30 $GHz$ FM2 &9.23 $    $ 9.26 $    $ 9.29&9.35 $    $ 9.29 $    $ 9.32&9.16 $    $ 9.10 $    $ 9.06&9.84 $    $ 9.77 $    $ 9.72\\
30 $GHz$ FM3 &9.34 $    $ 9.26 $    $ 9.13&9.04 $    $ 9.09 $    $ 9.02&9.03 $    $ 9.00 $    $ 8.88&8.88 $    $ 8.88 $    $ 8.72\\
44 $GHz$ FM1 &7.85 $    $ 7.60 $    $ 7.51&7.70 $    $ 7.50 $    $ 7.29&8.36 $    $ 8.23 $    $ 8.06&8.39 $    $ 8.32 $    $ 8.04\\
44 $GHz$ FM2 &8.57 $    $ 8.55 $    $ 8.44&7.87 $    $ 7.94 $    $ 7.92&8.76 $    $ 8.74 $    $ 8.58&8.55 $    $ 8.52 $    $ 8.35\\
44 $GHz$ FM3 &8.02 $    $ 7.93 $    $ 7.77&7.43 $    $ 7.38 $    $ 7.27&8.08 $    $ 8.02 $    $ 7.83&8.33 $    $ 8.31 $    $ 8.16\\
44 $GHz$ FM4 &7.87 $    $ 7.86 $    $ 7.75&8.35 $    $ 8.30 $    $ 8.04&7.72 $    $ 7.74 $    $ 7.60&8.51 $    $ 8.46 $    $ 8.22\\
\hline
\hline
\end{tabular}
\label{tab1}
\end{table}

\subsection{Equivalent noise temperature}

A method has been developed to achieve an accurate and unique equivalent noise temperature of the whole receiver. This method takes into account commercial noise sources, which have not a flat Excess Noise Ratio (ENR) versus frequency in millimetre-wave range, and $RF$ to $DC$ receiver performance along the band. Because the hot temperature of the used noise source and the BEM $RF$ gain show variations across the operating bandwidth, the next expression was used to obtain the global equivalent temperature ($T_{rec}$) \cite{aja03}:

\begin{equation}
T_{rec}=\frac{\sum^{f_{2}}_{f_{1}}{T_h {(f)}V_{det}{(f)}}-YT_c\sum^{f_{2}}_{f_{1}}{V_{det}{(f)}}}{(Y-1)\sum^{f_{2}}_{f_{1}}{V_{det}{(f)}}}	
\end{equation}

Where $T_h$ and $T_c$ are the hot and cold temperature of the noise source, $Y$ is the noise $Y-factor$, and $V_{det}$ is the detected voltage at each frequency when a CW signal, with a constant power sufficiently above white noise level, is applied at the BEM input.
The $Y-factor$ is given by:

\begin{equation}
Y={\frac{V_{det}|_{_h}}{V_{det}|_{_c}}}
\end{equation}
Where $V_{det}|_{_h}$ and $V_{det}|_{_c}$ are the receiver output voltages when two known source temperatures, hot and cold loads, are connected at the BEM input.

The $Y-factor$ was tested with a cold load and a hot load using a commercial noise source Q347B from Agilent. The equivalent noise temperature as a total power radiometer was slightly worse than on wafer measured noise figure of a naked MMIC due mainly to losses in the wave-guide to microstrip transition and to the readjustment of the bias point to decrease the ripple in the operating band, trading off noise temperature and effective bandwidth. This noise temperature has minimum impact on the global radiometer performance due to the high gain of the FEM.

Table \ref{tableTe} shows the values of the equivalent noise temperature for each flight model BEM at three different temperatures in the range of possible operating temperature. The large variability of the equivalent noise temperature of 44 GHz BEM units was due to their large dependence on the input matching network result, which was observed to be a very critical parameter, not easy to control during the assembly process of MMIC.

\begin{table}
\centering
\caption{Equivalent noise temperature of the BEM Flight models in Kelvin. (One unit of each band is a Flight Spare). Estimated error: $\pm$~20~$K$.}
\begin{tabular}{c|c|c|c|c}
\hline \hline
$$& Channel $A$ & Channel $B$ & Channel $C$ & Channel $D$ \\
\hline
$BEM$&$T_{low}$ $    $ $T_{nom}$ $    $ $T_{high}$&$T_{low}$ $    $ $T_{nom}$ $    $ $T_{high}$&$T_{low}$ $    $ $T_{nom}$ $    $ $T_{high}$&$T_{low}$ $    $ $T_{nom}$ $    $ $T_{high}$\\
\hline

30 $GHz$ FM1 &196 $    $ 317 $    $ 365&159 $    $ 294 $    $ 349&231 $    $ 349 $    $ 413&150 $    $ 292 $    $ 346 \\
30 $GHz$ FM2 &176 $    $ 288 $    $ 324&179 $    $ 299 $    $ 332&202 $    $ 316 $    $ 357&167 $    $ 307 $    $ 350 \\
30 $GHz$ FM3 &164 $    $ 286 $    $ 347&129 $    $ 272 $    $ 323&257 $    $ 342 $    $ 410&185 $    $ 301 $    $ 364 \\
44 $GHz$ FM1 &923 $    $ 856 $    $ 1006&734 $    $ 676 $    $ 745&798 $    $ 662 $    $ 744&	546 $    $ 643 $    $ 881 \\
44 $GHz$ FM2 &346 $    $ 494 $    $ 397&513 $    $ 674 $    $ 587&349 $    $ 426 $    $ 386&299 $    $ 405 $    $ 509 \\
44 $GHz$ FM3 &419 $    $ 467 $    $ 525&520 $    $ 595 $    $ 755&385 $    $ 437 $    $ 591&392 $    $ 437 $    $ 484 \\
44 $GHz$ FM4 &342 $    $ 561 $    $ 939&464 $    $ 459 $    $ 609&271 $    $ 380 $    $ 609&342 $    $ 510 $    $ 598 \\
\hline
\hline
\end{tabular}
\label{tableTe}
\end{table}

\subsection{Stability: 1/f noise}
The raw measurements of the output spectrum are used for the determination of the 1/f knee frequency. 
First, the output spectrum data are handled in a logarithmic scale in frequency. This way, it is possible to plot the spectral noise density composed by a white noise constant value, at high frequencies, plus the contribution of the 1/f noise at low frequencies. According to the definition, the 1/f noise knee frequency is the frequency at which the noise voltage spectrum density is at a level of $\sqrt{2}$ of the white noise voltage spectrum density level, for instance in $Vrms/sqrt(Hz)$. Using power density instead of voltage density then the level is 3 $dB$ above the white noise level (for instance in a scale of $dBm/Hz$).

\begin{figure}[t]
\centering
\includegraphics[width=.6\textwidth]{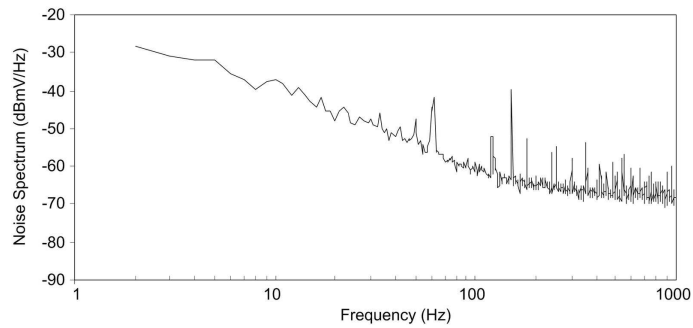}
\caption{ Typical low frequency noise power spectrum at the BEM output. Tested in a 50 Ohm spectrum analyser.}
\label{fig:fig19}
\end{figure}

The BEM low frequency power spectrum was characterized with a Hewlett Packard Vector Signal Analyzer HP81490A when a wave-guide matched load is connected to the input. The test was done at three temperatures: nominal (299 $K$), low (273 $K$) and high (326 $K$). The 1/f knee frequency was below 400 $Hz$ in all BEM units, much lower than the phase switching of the FEM (4096 $Hz$), so gain fluctuations of the back-end module did not impact on the global performance of the radiometer. The dominant 1/f noise source is attributed to the Schottky diode detector, since it refers directly to the diode current. The 1/f noise spectrum of each LNA alone was tested and the knee-frequency was about 13 $Hz$ for the N-ON LNA and about 15 $Hz$ for the N-OFF LNA. These results make evident that diode detector is mainly responsible for the knee-frequency of the BEM. The results for the four channels of a 30 $GHz$ BEM FM unit are given in Table \ref{table_fknee}. Figure \ref{fig:fig19} shows a typical noise spectrum of a flight model BEM at 30 $GHz$.

\begin{table}
\centering
\caption{1/f knee frequency (Hz) of 30 $GHz$ FM2 BEM unit.}
\begin{tabular}{c|c|c|c}
\hline \hline
 $Channel$ & $T_{low}$ & $T_{nom}$ & $T_{high}$ \\
\hline
$A$ & 75 & 90 & 103 \\
$B$ & 75 & 100 & 94   \\
$C$ & 74 & 100 & 117   \\
$D$ & 87 & 100 & 100  \\
\hline
\hline
\end{tabular}
\label{table_fknee}
\end{table}

\subsection{Linearity}

\begin{figure}[t]
\centering
\includegraphics[width=.6\textwidth]{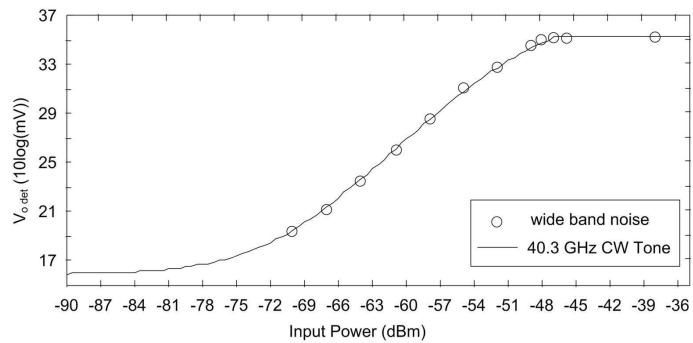}
\caption{44 $GHz$ QM BEM dynamic range.}
\label{fig:fig20}
\end{figure}

The detected voltage versus input power was measured in order to get the BEM dynamic range. A HP83650B generator was used as $CW$ source. Results for the 44 $GHz$ QM representative BEM are depicted in Figure \ref{fig:fig20} $($solid line$)$ for a 40.3 $GHz$ $CW$ signal input. In order to use a more realistic input signal, a wide band noise stimulus was used to measure the BEM sensitivity. This noise was obtained from a broadband white noise source, Q347B from Agilent, accordingly filtered and amplified by another 44 GHz BEM branch with only the radiofrequency chain (detector and DC amplifier not included). Results are also plotted in Figure \ref{fig:fig20} $($circles$)$. Taking into account the FEM gain and noise temperature, the BEM is always working under compression regime. The BEM non-linearity is a combination of the second MMIC LNA gain compression and the non-linearity of the diode detector. Given the dynamic range and the compression response of the BEM, it is not possible to identify the 1 dB compression point. Due to this compression effect, measured signal must be converted using the calibration curves, according to the procedures described in detail in in \cite{menella09}.

\begin{figure}[t]
\centering
\includegraphics[width=.6\textwidth]{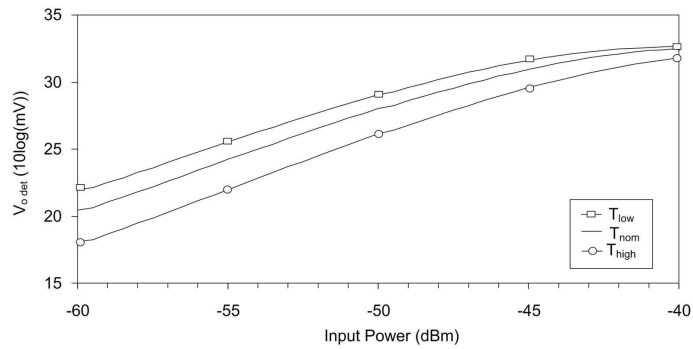}
\caption{Compression at 44 GHz (FM unit).}
\label{fig:fig21}
\end{figure}

Results of compression for channel A of a FM unit at 44 $GHz$ are shown in Figure \ref{fig:fig21}. The test was done at three temperatures: nominal, high and low. 

\section{Verification tests}

\subsection{Vibration and thermal vacuum}
Comprehensive vibration tests were performed, at different planes and frequency profiles, from 5 $Hz$ to 2000 $Hz$. The sequence applied in the QM BEM unit test was the following one for each axis: Low sine vibration, sine vibration, random vibration and low sine vibration surveys. The low level sinusoidal vibration surveys were conducted per each axis, prior and after performing sinusoidal and random vibration, at a sweep of 2 $oct/min$ and an acceleration of 0.5 $g$, in the range 5-2000 $Hz$, only one sweep per run. The sinusoidal vibration test consisted of a single sweep per axis, at a rate of 2 $oct/min$. In all vibration tests the BEM units were not operating.

\begin{figure}[t]
\centering
\includegraphics[width=.6\textwidth]{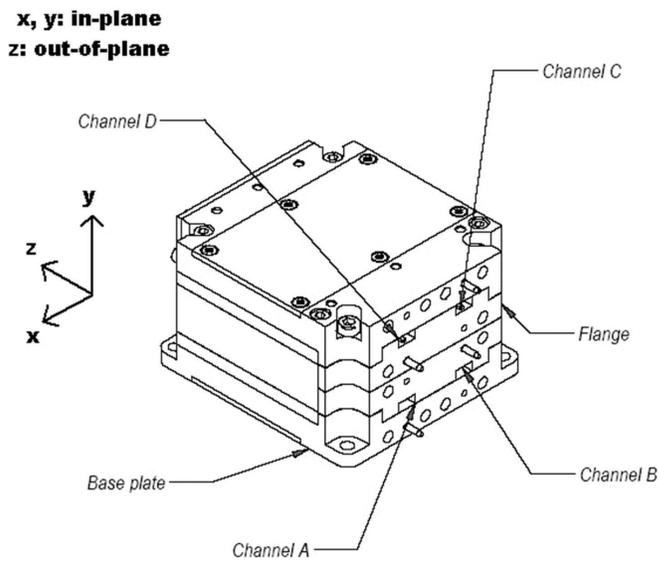}
\caption{BEM planes.}
\label{fig:fig22}
\end{figure}

The units were tested in 3 mutually perpendicular axes, 2 of them parallel to the base plate and the third one perpendicular to it. Figure \ref{fig:fig22} shows the axes orientation with relation to the BEM unit.

\begin{figure}[t]
\centering
\includegraphics[width=.4\textwidth]{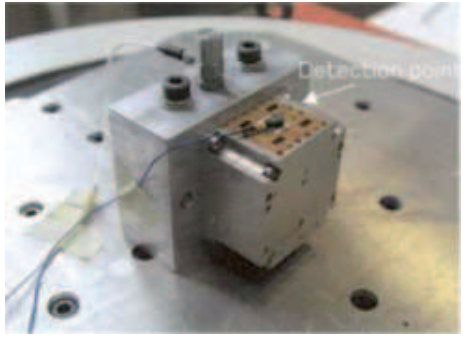}
\caption{30 GHz QM BEM attached to the shaker.}
\label{fig:fig23}
\end{figure}

As the specified vibration levels were defined with relation to the spacecraft axis system, a stiff fixture providing the right inclination for the BEM units was envisaged in order to match to the shaker axis. The BEM units were hard mounted on the fixture. This fixture guaranteed that the major modes of the BEM were not modified, in the sense that frequency shifts were below 5\% for lower frequency modes. The units were tested at the environmental temperature expected during satellite launch. A view of one BEM unit mounted on the shaker is in Figure \ref{fig:fig23}.

\begin{table}
\centering
 \caption{Acceptance sinusoidal vibration test levels.}
  \begin{tabular}{c|c|c}
\hline \hline
 $Axis$ & Frequency range $(Hz)$ & $Level$ \\
\hline
Out-of-plane $(OOP)$ & 5 - 21.25 & +/-8 mm \\
$$ & 21.25 - 100 & 20 g \\
\hline
In-plane $(IP-NFP, IP-PFP)$ & 5 - 21.25 & +/-8 mm \\
$$ & 21.25 - 100 & 16 g \\
\hline
\hline
\end{tabular}
\label{tablevibration}
\end{table}

For the FM units the acceptance sinusoidal vibration test consisted of a single sweep per axis, at a rate of $4$ oct/min. The acceptance levels applied in this test are indicated in Table \ref{tablevibration}. The acceptance random vibration test was done according to the levels and durations presented in Table \ref{tablerandom}. 

\begin{table}
\centering
\caption{Acceptance random vibration test levels.}
  \begin{tabular}{c|c|c|c|c}
\hline \hline
 $Axis$ & Frequency $(Hz)$ & Power Spectral Density $(g^{2}/Hz)$ &	$G_{rms}$	& Time per axis $(s)$ \\
\hline
$$ & 20 &	0.0482 & $$ & $$ \\
$$ & 80	& 0.192 & $$ & $$ \\
$Out-of-plane $ & 220 &	6.912 & $$ & $$ \\
$(OOP)$ & 240 &	6.912 & 25.05 & 60 \\
$$ & 300 &	0.832 & $$ & $$ \\
$$ & 500 &	0.832 & $$ & $$ \\
$$ & 2000 &	0.00822 & $$ & $$ \\
\hline
$$ & 20 &	0.0241 & $$ & $$ \\
$$ & 80 &	0.096 & $$ & $$ \\
$In-plane $ & 150 &	3.456 & $$ & $$ \\
$(IP-NFP, IP-PFP)$ & 170 &	3.456 & 20.01 & 60 \\
$$ & 300 &	0.704 & $$ & $$ \\
$$ & 500 &	0.704 & $$ & $$ \\
$$ & 2000 &	0.00411 & $$ & $$ \\
\hline 
\hline
\end{tabular}
\label{tablerandom}
\end{table}

The proper BEM performance, after the vibration of each unit, was checked by post-dynamic performance verification tests. These checks were performed by measuring a small number of indicative requirements of the BEM, like power consumption, in order to be sure that the unit was still alive and performing well. The electrical behaviour, in terms of noise and RF to DC conversion, did not change after vibration and thermal vacuum tests. The tested deviations were within the range of the test equipment measurement accuracy.

\begin{figure}[t]
\centering
\includegraphics[width=.6\textwidth]{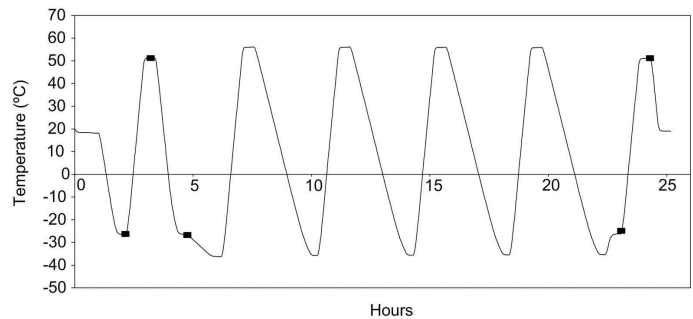}
\caption{Thermal profile for a FM unit test (RPT instants highlighted with squares).}
\label{fig:fig24}
\end{figure}

Thermal vacuum tests for BEM flight units were done basically by six thermal cycles, with a total duration of about 25 hours, between $-35^{\circ}C$ and $+55^{\circ}C$. The thermal profile is depicted in Figure \ref{fig:fig24}. During thermal cycling, a set of Reduced Performance Tests (RPT) have been performed. As in the vibration tests, after thermal vacuum test the survival condition and functionality of each unit was checked. 

Each  unit  in turn was screwed  to the base plate of the vacuum chamber and low thermal resistance ensured by thermal silcone compound. Temperature cycling in the acceptance range was thus achieved by changing the base plate temperature. The DC feedthroughs of the chamber were connected to the BEM, in order to have available the DC power lines outside the chamber. Output channel signals and DC consumption were continuously monitored by the Agilent 34970A Data Logger with the 34901A switching unit.

\begin{figure}[t]
\centering
\includegraphics[width=.4\textwidth]{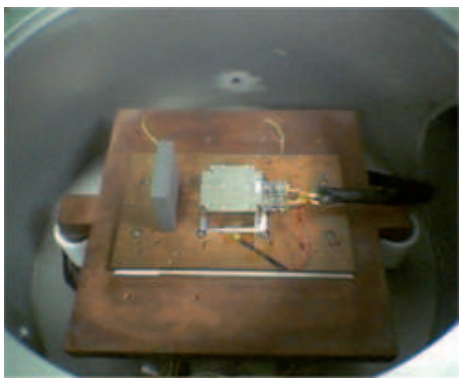}
\caption{BEM fixation to base plate of thermal vacuum chamber.}
\label{fig:fig25}
\end{figure}

As it is shown in Figure \ref{fig:fig25}, the BEM waveguide inputs were left open. In order to avoid the detection of any noise signal, an aluminium wall, covered with a layer of microwave absorber, has been located at a distance of 2.5 $cm$ from the BEM.

\subsection{Electromagnetic compatibility}

Very strict EMC tests were performed on BEM units, according to specifications. Both conducted and radiated emission and susceptibility tests were made. Interfering signals covering the range from 30 $Hz$ to 18 $GHz$ depending on the individual test were used. Radiated tests included electric and magnetic fields emission and susceptibility. In the case of conducted EMC tests the emission and the susceptibility were tested through the power supply lines $\pm$ 5 $Volt$. The most difficult test to fulfil was the conducted susceptibility, because the BEM units do not have DC to DC converters inside, and power supply fluctuations appeared at the BEM detected output. Special filtering on the power supply input lines was used to avoid susceptibility at low frequencies.

\section{Conclusions}

Back End Modules at 30 and 44 $GHz$ for Planck Low Frequency Instrument have been designed, manufactured and tested. They have successfully fulfilled the electrical, mechanical and environmental requirements of Planck satellite mission. Qualification model units and Flight Model units have demonstrated the compliance with the required performances.

%________________________________________________________________
\acknowledgments
This work has been supported by the Spanish "Plan Nacional de I+D+i", Programa Nacional de Espacio, grants ESP2002-04141-C03-01/02/03 and ESP2004-07067-C03-02. The authors would like to thank Eva Cuerno and Alexandrina Pana for the assembly of the BEM prototypes. Planck is a project of the European Space Agency with instruments funded by ESA member states, and with special contributions from Denmark and NASA (USA). The Planck-LFI project is developed by an International Consortium lead by Italy and involving Canada, Finland, Germany, Norway, Spain, Switzerland, UK, USA.

\end{document}